\documentclass[twocolumn,floatfix,prb,aps]{revtex4-1}
\usepackage{color,amsmath,amssymb,amsfonts,graphicx,tabularx}
\usepackage{epstopdf}
\usepackage{amsmath}  
\usepackage{amsfonts} 
\usepackage{graphicx} 
\usepackage[unicode=true,colorlinks=true]{hyperref}
\hypersetup{linkcolor=blue,citecolor=blue,urlcolor=blue}

\makeatletter
\newcommand*{\rom}[1]{\expandafter\@slowromancap\romannumeral #1@}
\makeatother

\newcommand{\be}{\begin{equation}}
\newcommand{\ee}{\end{equation}}

\newcommand{\ba}{\begin{eqnarray}}
\newcommand{\ea}{\end{eqnarray}}

\begin{document}
\title{Mean field approximations for short range four body interactions at $\nu=3/5$}
\author{Bartosz Ku\'{s}mierz $^{1,2}$, Arkadiusz W\'{o}js$^1$, Sreejith G J$^2$}
\affiliation{$^1$ Department of Theoretical Physics, Wroclaw University of Science and Technology, Poland\\ $^2$ IISER Pune, Dr Homi Bhabha Road, Pune 411008, India}
\begin{abstract} 
Trial wavefunctions like the Moore-Read and Read-Rezayi states which minimize short range multibody interactions are candidate states for describing the fractional quantum Hall effects at filling factors $\nu = 1/2$ and $\nu = 2/5$ in the second Landau level. These trial wavefunctions are unique zero energy states of three body and four body interaction Hamiltonians respectively but are not close to the ground states of the Coulomb interaction. Previous studies using extensive parameter scans have found  optimal two body interactions that produce states close to these. Here we focus on short ranged four body interaction and study two mean field approximations that reduce the four body interactions to two body interactions by replacing composite operators with their incompressible ground state expectation values. We present the results for pseudopotentials of these approximate interactions. Comparison of finite system spectra of the four body and the approximate interactions at filling fraction $\nu=3/5$ show that these approximations produce good effective descriptions of the low energy structure of the four body ineraction Hamiltonian. The approach also independently reproduces the optimal two body interaction inferred from parameter scans. We also show that for $n=3$, but not for $n=4$, the mean field approximations of the $n$-body interaction is equivalent to particle hole symmetrization of the interaction.
\end{abstract}
\maketitle

\section{Introduction}
The physics of electrons confined to two dimensions in the limit of high magnetic fields, is described by a Hamiltonian that contains no kinetic energy but only the Coulomb interaction term, with the kinetic energy indirectly manifesting itself through the holomorphic nature of the Hilbert space. Interacting electrons in this Hilbert space exhibit a rich set of  topological and conventionally ordered phases \cite{Tsui1982,WignerCrystal2016,fogler2002stripe}. The phases in the lowest Landau level can be explained accurately using composite fermion wavefunctions\cite{Laughlin1983,Jain1989}. Structure of fractional quantum Hall effect in the second Landau level has been harder to explain using variational studies. Among the several candidate wavefunctions proposed to describe these states are a set of clustered states including the Pfaffian and the $k=3$ Read Rezayi states occurring at filling fractions $1/2$ and $3/5$ in this Landau level.\cite{moore1991nonabelions,RRState1999} The correlations contained in these states are such that they minimize certain model Hamiltonians \cite{RRState1999,Simon07} that penalize specific shortrange configurations of clusters of few particles. These wavefunctions do not have large overlaps with the physical two body Coulomb interaction ground states. Moore Read state, for instance, has an overlap of $0.69$ with the Coulomb ground state in a system of size $N=14$.\cite{Sreejith11a} However, it has been argued that they capture the topological properties of the Coulomb ground states. 

Since these states minimize a model interaction energy rather than the physical two body Coulomb interaction energy, it is interesting to ask whether there is a two body interaction that produce ground states that are close to these clustered states. One approach to addressing this is to consider general short range two body interactions parametrized by Haldane psuedopotentials and scan the parameter space to identify the optimal pseudopotentials that produce a homogeneous ground state with maximal overlap with the clustered state.\cite{Kusmierz18a,kusmierz2016jack} Another approach suggested in Ref \onlinecite{Sreejith17} is to make use of the model $n$-body interactions that annihiliate these clustered state to arrive at apprximate two body body interactions via a mean field mapping. These two approaches surprisingly produce the same optimal interaction in the case of the Moore Read state which is annihilated by the three body interaction. \cite{Kusmierz18a}

In this study, we explore an extension of the mean field approximation to the case of the four body interaction that produces the $k=3$ Read-Rezayi state as the ground state. There are two possible ways to map the four body interaction to a two body interaction - (1) by replacing two pairs of composite operators $c_i^\dagger c_j$ or (2) by replacing a $c_i^\dagger c_j^\dagger c_k c_l$ with their ground state expectation values. The former method is scalable to larger system sizes allowing us to extrapolate to the mean field two body psuedopotentials in the thermodynamic limit. Interestingly, the mean field interaction matches exactly with what was obtained through extensive parameter scan in Ref-[\onlinecite{Kusmierz18a}]. Note that, in addition to the interactions whose influence we explore in this study, the state describing the physical system can be qualitatively changed by the presence of disorder especially at filling fraction $\nu=\frac{1}{2}$.\cite{PhysRevLett.117.096802,PhysRevB.98.045112,PhysRevLett.121.026801} We however consider spin polarized systems which are disorder free for tractability using finite system studies.

The general $n>2$ body interaction, and in particular the three body interaction is not particle hole symmetric. Particle-hole symmetrization of the short range three body interaction Hamiltonian produces a two body interaction which has a low energy spectrum close to that of the original three body interaction.\cite{Peterson08,PetersonScarola19} It was found in Ref-[\onlinecite{Sreejith17}] that the mean field two body approximation reproduced the same two body interaction as the symmetrization. We explain why the two interactions exactly reproduce the same spectra, and extend this analysis to the case of the four body interaction and show that this exact relation between symmetrization and mean field approximation is restricted to the case of the three body interaction. 

In Sec \ref{MF:desc}, we introduce the notion of Haldane pseudopotentials for general $n$-body interactions, followed by a description of the mean field approximation, discussing the idea for the case of three body and four body interactions. The mean field approximation of the four body interaction can be defined to produce a three body or a two body interaction. The latter can be arrived at in two different ways. Every method results in a rotationally symmetric interaction and therefore can be specified in terms of the pseudopotentials. In Sec-\ref{sec:pseudo}, we present the results of the psuedopotentials of the mean field two and three body interactions for finite systems as well as in the large system limit.
In Sec \ref{sec:PHSym}, we discuss the relation between the mean field approximation and the particle hole symmetrization / antisymmetrization of the interactions. We show that the mean field approximation to the three body interaction and symmetrization produce the same spectrum. We show that this result does not generalize to the case of the four body interaction. 
In Sec-\ref{sec:numtests}, we compare the finite system spectra of the mean field interactions with that of the exact four body ineraction. The results suggest that the mean field interaction closely reproduces the effective physics of the incompressible and few quasiparticle/quasihole systems. Approximate formulae for the mean field interaction pseudopotentials that could be used in further numerical studies are presented in the appendix.

\section{The types of the MF in the spherical geometry}\label{MF:desc}
  
We use the standard  Haldane  spherical geometry \cite{Haldane83,Fano86}, in which $N$ electrons are confined to the surface  of  a  sphere  of  radius $R$,  with a uniform, perpendicular magnetic field $B$ being provided by a Dirac magnetic monopole of strength $2 Q \phi_0$ ($2Q$ is an integer) placed at the center of the sphere, where the flux quanta $\phi_0$ is $h c/e$. The corresponding magnetic length has a value $\ell_B =R/\sqrt{Q}$. The $N$-electron Hilbert space is spanned by the configurations $\left | \bf p\right \rangle = |p_1,p_2,...,p_N \rangle$ of electrons occupying orbitals $p_i$ (with $p_i\in\{-Q,-Q+1,\dots Q\}$ ). The general four body Hamiltonian can be written as 
\begin{equation} \label{eq:general_ham} 
\mathcal{H}^{(4)} = \sum_{p_i;q_i} V^{(4)}_{{\bf q};{\bf k}} c^\dagger_{p_4} c^\dagger_{p_3} c^\dagger_{p_2} c^\dagger_{p_1} c_{q_1} c_{q_2} c_{q_3} c_{q_4},
\end{equation}
where ${\bf p} =(p_1,p_2,p_3,p_4),{\bf q}= (q_1,q_2,q_3,q_4)$ and the indices correspond to an ordered set of $L_z$ quantum numbers of the electrons ($p_i< p_{i+1}$ and $q_i< q_{i+1}$). 

When considered system has additional symmetries, certain constraints can be imposed reducing the number of independent parameters that describe the interaction. A rotationally symmetric four body interaction of spinless fermions on the Haldane sphere can be described by a sequence of generalized Haldane four body pseudopotentials $\{V_l^{(4)}\}_l$, where $V_l$ is the an energy needed for four particles to be in a relative angular momentum of $4Q-l$. Thus the Hamiltonian has the form 
\begin{equation}
{\mathcal H}^{(4)} = \sum_{l=6}^{2Q} V^{(4)}_l P_{4Q-l}^{(4)},
\end{equation}
where $P_{4Q-l}^{(4)}$ is a projector on to the relative angular momentum $4Q-l$ subspace of four particles. This can be explicitly written as follows:
\begin{equation}
P_{L}^{(4)} = \sum_{a}\sum_{Lz=-L}^L \sum_{{\bf p},{\bf q}} \psi_{L,Lz,a} ({\bf p}) \psi_{L,Lz} ({\bf q}) \prod_{i=1}^4 c_{p_i}^\dagger \prod_{i=1}^4 c_{q_i}
\end{equation}
where $\psi_{L,Lz,a}(p_1,p_2,p_3,p_4)$ are the Clebsch Gordan coefficients when the four particle state of total angular momentum $L$ and $z$-component angular momentum $L_z$ are expanded in the single particle basis $\left|p_1,p_2,p_3,p_4\right\rangle$ ie 
\begin{equation}
\left |{\psi_{L,Lz,a}}\right \rangle = \sum_{\bf p}\psi_{L,Lz,a}(p_1,p_2,p_3,p_4) \left|p_1,p_2,p_3,p_4\right\rangle
\end{equation}
Pauli exclusion implies that the allowed values of $l=4Q-L$ for splinless fermions are $l=6,8,9,...$. The index $a$ corresponds to the possibility of different multiplets of angular momentum $L$. Short range interaction corresponds to smaller values of $l$. Several independent angular momentum multiplets can occur for $l>9$ and the additional quantum number $a$ is then required \cite{Simon07}. However in this article, for simplicity, we shall consider only those interactions for which $l=6$ or $l=8$ and the quantum nunber $a$ will not be required.



The wave functions $\psi^{L,Lz}_{\bf p}$ can be obtained via exact diagonalization of four particles in a generic rotationally symmetric 2-body interaction.


In Ref [\onlinecite{Sreejith17}], the authors introduced a mean field mapping of the three body interaction to a two body interaction by replacing a single quadratic composite operator $c^\dagger_{q_1}c_{k_1}$ with its expectation value. Analysis was restricted to states in the vicinity of the incompressible ground states, for which the expectation values take a simple form $\left\langle c^\dagger_{q_1}c_{k_1} \right \rangle = \nu\,\delta_{q_1,k_1}$ because of angular momentum conservation. We can apply this method here, resulting in reduction of the four body interaction to a three body interaction.

The general four body Hamiltonian (Eq \ref{eq:general_ham}) can be written without the restriction on the ordering of the single particle angular momenta as 
\begin{equation}
{\mathcal H}^{(4)} = \sum_{p_i,q_i=-Q}^Q c^\dagger_{p_4} c^\dagger_{p_3} c^\dagger_{p_2} c^\dagger_{p_1} \frac{V^{(4)}_{{\bf p},{\bf q}}}{4! 4!}
c_{q_1} c_{q_2} c_{q_3} c_{q_4}\nonumber.
\end{equation}
Antisymmetry of $V_{{\bf p},{\bf q}}$ is assumed under interachange of single particle indices within ${\bf p}$ and ${\bf q}$. Upon applying the mean field approximation we obtain a three body interaction of the form.
\begin{equation}
{\mathcal H}^{(3)} = \sum_{p_i,q_i=-Q}^Q c^\dagger_{p_3} c^\dagger_{p_2} c^\dagger_{p_1} \frac{V^{(3)}_{p_1,p_2,p_3;q_1,q_2,q_3}}{3!3!} c_{q_1}c_{q_2}c_{q_3},
\end{equation}
where $V^{(3)}$ is given by the partial trace over one pair of indices
\begin{equation} \label{eq:mf_v3}
V^{(3)}_{p_1,p_2,p_3;q_1,q_2,q_3} = \nu
\sum_{p_4,q_4=-Q}^{Q} \delta_{p_4q_4} V^{(4)}_{{\bf p},{\bf q}}.
\end{equation}
The three body pseudopotentials $V^{(3)}_l$ of the mean field three body Hamiltonian can be obtained by numerically diagonalizing a system of three particles. The energy of the three particle cluster of angular momentum $4Q-l$ gives the psuedopotential $V_l^{(3)}$.

A mean field approximation of a similar kind applied now to the above three body Hamiltonian results in a two body interaction. The two body pseudopotentials $V^{(2)}_l$ can now be obtained by diagonalizing a two particle system. Thus reduction (four to two body) is obtained by applying ``single'' mean field approximation twice. Since we approximated operators $c^{\dagger}c$ with expected value $\nu \delta_{q_1,k_1}$, we will denote this type of mean field reduction by $MF^2_{\langle c^{\dagger}c\rangle}$. The intermidiate step of reduction of four body Hamiltonian to three body by applying the approximation once is denoted $MF_{\langle c^{\dagger}c\rangle}$.

One can construct an alternative mean field reduction of four to two body Hamiltonians by replacing the composite quartic operator $c^\dagger_{q_2} c^\dagger_{q_1} c_{k_1} c_{k_2}$ with its exceptation value in the incompressible ground state. Such an expectation value is not easy to calculate, even when one considers the homogeneous ground state. So we approach the problem with numerical calculations of correlations in the ground state. For each pair of indexes $(q_1,q_2)$ we calculate expected value of $c^\dagger_{q_2} c^\dagger_{q_1} c_{k_1} c_{k_2}$, which is later used to infer mean field two body Hamiltonian. We will denote this mean field mapping by $MF_{\langle c^{\dagger}c^{\dagger}c c\rangle}$. In this study, we will use the $3/5$ filling fraction to explore the mean field approximation, as an incompressible state ($k=3$ Read-Rezayi state) is produced by the short range four body interaction at this filling fraction.\cite{Read99}

\begin{figure} 
\includegraphics[width=.45\columnwidth]{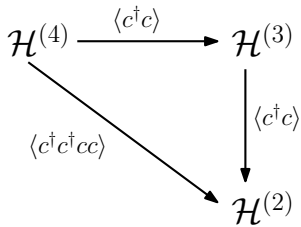}
\caption{The schematic diagram two possible methods of MF reduction of four body interaction to two body interaction. The $MF^2_{\langle c^{\dagger} c\rangle}$ method  replaces a pair of creation and annihilation operators with their expectation value twice. As an intermediate step one obtains three body operator ${\mathcal H}^{(3)}$. The diagonal arrow corresponds to the mean field mapping using numerically obtained correlations from Read-Rezayi state $c^\dagger_{q_2} c^\dagger_{q_1} c_{k_1} c_{k_2}$. \label{diag1}}
\end{figure}

The mean field Hamiltonian needs to be rotationally symmetric in order to be able to calculate the pseudopotentials. It can be easily seen that the methods produce rotationally symmetric approximations. Due to rotational symmetry of the original four body interaction, the interaction parameters $V_{{\bf p}{\bf q}}$ are elements of a linear combination of projections onto angular momentum subspaces. Therefore, these interaction parameters form a rotationally invariant tensor. The mean field approximations $MF_{\langle c^{\dagger}c\rangle}$ and $MF_{\langle c^{\dagger}c\rangle}^2$ correspond to contraction of indices of this tensor with the indices of the rotationally invariant tensors $\delta_{p_4q_4}$ (Eq \ref{eq:mf_v3}) and $\delta_{p_4q_4}\delta_{p_3q_3}$ respectively. Therefore the mean field interaction parameters $V^{(3)}_{p_1p_2p_3;q_1 q_2 q_3}$ and $V^{(2)}_{p_1p_2;q_1 q_2}$ obtained this way are rotationally invariant. Rotational invaraince implies that the  interaction parameters of the mean field Hamiltonian are linear combinations of angular momentum projection operators, the coefficients of which give the pseudopotentials. In the case of $MF_{\langle c^{\dagger}c^\dagger cc\rangle}$, $V_{\bf{p}\bf{q}}$ is contracted with the correlation $\langle c_{p_1}^{\dagger}c_{p_2}^\dagger c_{q_1}c_{q_2}\rangle$ which is again rotationally symmetric due to the rotational symmetry ($L=0$) of the ground state. 
The information contained in the correlation function can indeed be represented as linear combinations of two particle angular momentum projection operators (Such expansions for specific finite systems are presented in Appendix \ref{App:4fermioncorrel})

\section{Pseudopotentials for the mean field mapped interactions}
\label{sec:pseudo}
In this section we apply the mean field mapping to the specific cases and present the results for the psuedopotentials calculated from the different mean field mappings. In addition to the  the short range four body interaction ($V_6=1,V_{l\neq 6}=0$), we also consider the case of the longer range four body interaction ($V_8=1,V_{l\neq 8}=0$). The latter is a hollow core four body interaction. Analogous hollow core two and three body interactions have been found to be useful in studies of FQH states such as at $\nu=4/11$.\cite{PhysRevLett.105.196801,PhysRevLett.112.016801,PhysRevB.71.245331}
\subsection{$MF^2_{\langle c^{\dagger}c\rangle}$}\label{deltadelta}
As described before, $MF_{\langle c^{\dagger}c\rangle}$ applied twice ($MF^2_{\langle c^{\dagger}c\rangle}$) maps the four body interaction to a two body interaction. Two body pseudopotentials are extraced using a direct diagonalization of system of only two particles. Since Hilbert space for such systems is relatively small, it is possible to calculate coefficients for systems with large $2Q$. In Table \ref{tab:mfcc} we present values of the pseudopotentials for the two largest systems that we have studied. Irrespective of system size, only the first three allowed two body pseudopotentials are non-zero in the mean field mapping of the  $V_6=1,V_8=0$ interaction and only the first four allowed two body pseudopotentials are found to be non-zero in the mean field mapping of the $V_6=0,V_8=1$ interaction.
\begin{table}[!h]
\centering
\begin{tabular}{|c|c|c||c|c|}
\hline 
 & \multicolumn{2}{c||}{$V_{6}^{(4)}=1,V_{8}^{(4)}=0$} & \multicolumn{2}{c|}{$V_{6}^{(4)}=0,V_{8}^{(4)}=1$}\tabularnewline
\hline 
$V^{(2)}_{n}\downarrow$ & $2Q=60$ & $2Q=62$ & $2Q=60$ & $2Q=62$\tabularnewline
\hline 
$V_{1}$ & $7.01312$ & $7.02084$ & $6.14707$ & $6.15839$\tabularnewline
\hline 
$V_{3}$ & $3.46795$ & $3.46901$ & $2.68532$ & $2.68741$\tabularnewline
\hline 
$V_{5}$ & $1.26689$ & $1.26630$ & $1.04738$ & $1.04576$\tabularnewline
\hline 
$V_{7}$ & 0 & 0 & $1.81347$ & $1.81125$\tabularnewline
\hline 
\end{tabular}
\caption{Two body pesudopotentials obtained by mean field mapping $MF^2_{\langle c^{\dagger}c\rangle}$ of the four body interactions. Data is presented for the two largest systems studied, and for the two types of interactions first a short-range repulsion where a single four body pseudopotential $V_6$ is nonzero and second a longer range interaction with only $V_8$ being nonzero.
}\label{tab:mfcc}
\end{table}

\begin{figure} 
  \begin{center}      
     \scalebox{0.6}{\input{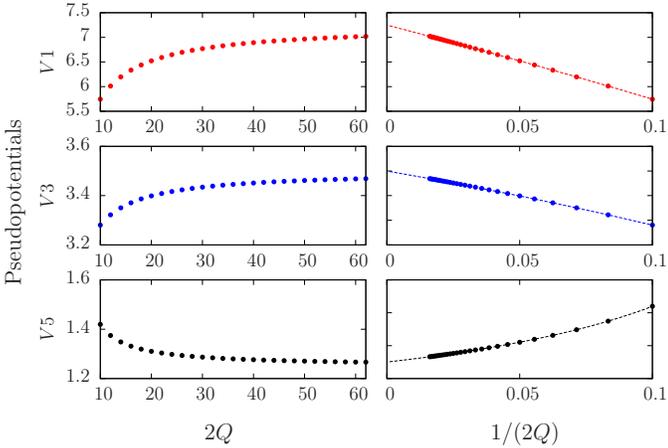}}                           
     \caption{(left): Two body pesudopotentials obtained by the mapping $MF^2_{\langle c^{\dagger}c\rangle}$ of four body short-range repulsion with $V_6=1, V_{8}=0$. (right): Same information shown as a function $1/2Q$ to show convergence to the values in the $2Q\to\infty$ limit. The dotted lines indicate the fitting function $a+b/(c-2Q)$}\label{fig:mfcc_ps6}
   \end{center}
\end{figure}
      
\begin{figure} 
  \begin{center} 
     \scalebox{0.6}{\input{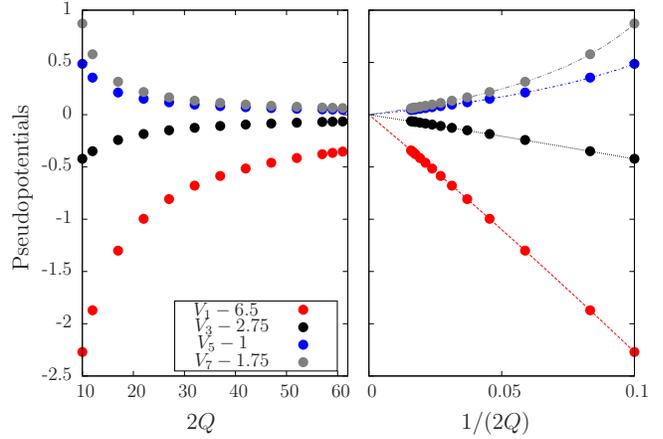}}
     \caption{(left) Two body pesudopotentials obtained by the mapping $MF^2_{\langle c^{\dagger}c\rangle}$ of longer range four body repulsion ($V_8=1, V_6=0$) (right) Same information plotted as a function of $1/2Q$. The vertical axis shows the deviation from the values in the $2Q\to \infty$ limit. The dotted lines indicate the fitting function $a+b/(c-2Q)$} \label{fig:mfcc_ps8}
   \end{center} 
\end{figure}

The values of the two body pseudopotentials for smaller systems are presented in the Fig \ref{fig:mfcc_ps6} and Fig \ref{fig:mfcc_ps8}. The data allows an extrapolation to the $2Q\to\infty$ limit using a simple function $V(2Q) = a+\frac{b}{c-2Q}$. Uncertainties of the coefficients $a,b,c$ are very small; we present them in the Table \ref{tab:coef} in the Appendix \ref{app1}. For the short-range four body repulsion the two body pseudopotentials in the $2Q\to\infty$ limit have values $V_1: V_3:V_5 = 7.24975: 3.50013:  1.24996 \approx  5.8:2.8:1$. This mean field two body interaction is identical to the optimal two body interaction for the $k=3$ Read-Rezayi state obtained in Ref [\onlinecite{Kusmierz18a}], wherein the authors had studied systems of sizes upto $N=21$ and $2Q=32$ and found the ratio to be $6:3:1$. At the same flux, the mean field approximation gives pseudopotentials in the ratio $5.3:2.6:1$.
When comparing the pseudopotentials, we note that the numerical search for the optimal interaction (Ref [\onlinecite{Kusmierz18a}]) was performed on a finite grid in the paremeter space, which is expected to result in finite errorbars on the optimal pseudopotentials. Similar mean field approximation to the $V_8 =1$ interaction gives the ratios $V_1: V_3:V_5:V_7 \approx 6.5: 2.75:  1: 1.75$ in the large $2Q$ limit (Fig \ref{fig:mfcc_ps8}).

Linearity of the the mean field mapping ${\rm MF}(H_1)+{\rm MF}(H_2)={\rm MF}(H_1+H_2)$ implies that the mean field pseudopotentials of a four body interaction with $V_6=A,V_8=B$ can be obtained as the corresponding linear combination of the mean field psuedopotentials of $V_6=1,V_8=0$ and $V_8=1,V_6=0$ given in the tables before.

\subsection{$MF_{\langle c^{\dagger}c\rangle}$}

When the mean field mapping $MF_{\langle c^{\dagger}c\rangle}$ is applied to a four body interaction only once, we  obtain a three body interaction. The three body pseudopotentials obtained by diagonalizing a system of three particles is presented in the Table \ref{tab:mfcc3b}.
\begin{table}[!h]
\centering
\begin{tabular}{|c|c|c||c|c|}
\hline 
 & \multicolumn{2}{c||}{$V_{6}^{(4)}=1,V_{8}^{(4)}=0$} & \multicolumn{2}{c|}{$V_{6}^{(4)}=0,V_{8}^{(4)}=1$}\tabularnewline
\hline 
$V_{n}^{(3)}\downarrow$ & $2Q=47$ & $2Q=52$ & $2Q=47$ & $2Q=52$\tabularnewline
\hline 
$V_{3}$ & $3.10234$ & $3.10812$ & $1.9913$ & $2.0027$\tabularnewline
\hline 
$V_{5}$ & $1.19654$ & $1.19537$ & $0.6742$ & $0.6726$\tabularnewline
\hline 
$V_{6}$ & $0.98568$ & $0.98589$ & $0.0783$ & $0.0770$\tabularnewline
\hline 
$V_{7}$ & 0 & 0 & $1.6232$ & $1.6187$\tabularnewline
\hline 
$V_{8}$ & 0 & 0 & $0.9067$ & $0.9084$\tabularnewline
\hline 
\end{tabular}
\caption{ Three body pesudopotentials obtained by reduction of four body interactions to three body using the mapping $MF_{\langle c^{\dagger}c\rangle}$. Data are presented for the two largest systems and for the two types of interactions - short-range repulsion ($V_6=1,V_8=0$) and the longer range repulsion ($V_6=0,V_8=1$) 
}\label{tab:mfcc3b}
\end{table}

\begin{figure} 
  \begin{center}      
     \scalebox{0.60}{\input{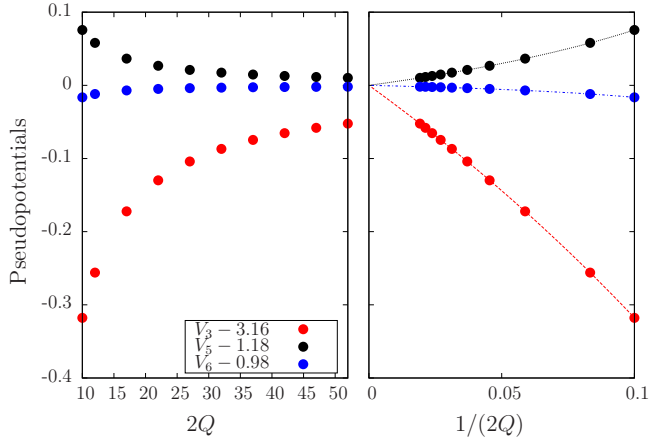}}                           
     \caption{Three body pesudopotentials obtained by $MF_{\langle c^{\dagger}c\rangle}$ mapping of the four body short-range repulsion ($V_6=1, V_{8}=0$) to three body interaction. The dotted lines indicate the fitting function $a+b/(c-2Q)$} \label{fig:mfcc_ps6_3} 
   \end{center} 
\end{figure}

\begin{figure} 
  \begin{center}      
     \scalebox{0.60}{\input{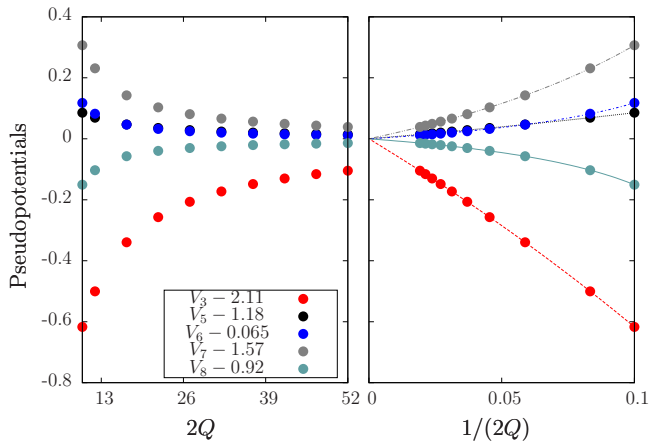}}                           
     \caption{Three body pesudopotentials obtained by $MF_{\langle c^{\dagger}c\rangle}$ - reduction of longer range four body interaction ($V_8=1, V_6=0$) to three body interaction. The dotted lines indicate the fitting function $a+b/(c-2Q)$} \label{fig:mfcc_ps8_3} 
   \end{center} 
\end{figure}

The values of the three body pseudopotentials in the large $2Q$ limit can also be inferred using a fitting function $a+\frac{b}{c-2Q}$ (Fig \ref{fig:mfcc_ps6_3} and Fig \ref{fig:mfcc_ps8_3}). For the values of the coefficients $a,b,c$ and their dispersion see tab. \ref{tab:coef}.

\subsection{$MF_{\langle c^{\dagger}c^{\dagger} c c\rangle}$}

As described in Sec-\ref{MF:desc}, one can directly map a four body to an approximate two body interaction by replacing the composite quartic operator $c_i^\dagger c_j^\dagger c_k c_l$ by the ground state expectation values. Unlike the previous two cases, where the only information required to define the mapping came from assumption of homogeneiety and rotational symmetry of the ground state (which implied $\langle c_i^\dagger c_k\rangle\propto \delta_{ik}$), definition of the $MF_{\langle c^{\dagger}c^{\dagger} c c\rangle}$ mapping requires more specific knowledge of the many body state in which $\langle c_i^\dagger c_j^\dagger c_k c_l \rangle$ is calculated. This prevents us from implementing and exploring this mean field calculation for systems larger than $2Q=27$.

\begin{table}[!h]
\centering
\begin{tabular}{|c||c|c|}
\hline
 \multicolumn{3}{|c|} {4-body $V_6=1$; Corr $V_6=1$
} \\ \hline
$V_n^{(2)}$ & $2Q=22$ & $2Q=27$   \\
\hline
  $V_1$ & 2.48198   &  2.45339             \\ \hline
  $V_3$ & 1.25684   &  1.22514           	\\ \hline
  $V_5$ & 0.43826  	&  0.41369  		    \\ \hline 

\end{tabular}
\caption{Two body pesudopotentials from the mapping $MF_{\langle c^{\dagger}c^{\dagger}cc\rangle}$of the short range four body pseudopotential ($V_6=1,V_8=0$). The correlation $\langle c^{\dagger}c^{\dagger}cc\rangle$ are taken from the Read-Rezayi state.
}
\label{tab:cccc6}
\end{table}

We estimated $\langle c_i^\dagger c_j^\dagger c_k c_l \rangle$ in the incompressible ground state ($k=3$ Read-Rezayi state) of the short range four body interaction ($V_6 = 1, V_{8} =0$) of $N=15$ and $N=18$ particles at flux $2Q=22$ and $2Q=27$ respectively. From the two body interaction obtained from this approximation, the pseudopotential can again be estimated from the energies of two particles. Table \ref{tab:cccc6} contains pseudopotentials of reduced interaction for the largest systems that we studied. The pseudopotentials at $2Q=27$ occur in the ratio $V_1:V_3:V_5=5.8:3.0:1$ matching closely with results of Ref \onlinecite{Kusmierz18a}

The longer range four body interaction ($V_8 = 1, V_{6} =0$) does not produce an incompressible state at $2Q=\frac{5}{3}N-3$ at every $N$. In the absence of a gapped ground state, it is not clear that such a mean field approximation will work. 
Nevertheless, a mean field approximation can still be constructed for the $V_8=1$ interaction using the correlations calculated from its ground state in the $L=0$ sector. Table \ref{tab:cccc8} presents two body pseudopotentials of mean field reduction of this interaction.

\begin{table}[!h]
\centering
\begin{tabular}{|c||c|c|}
\hline
 \multicolumn{3}{|c|} {4-body $V_8=1$; Corr $V_8=1$
} \\ \hline
$V_n^{(2)}$ & $2Q=22$ & $2Q=27$   \\
\hline
  $V_1$ & 2.00931   &  1.91807             \\ \hline
  $V_3$ & 1.12664   &  1.22146           	\\ \hline
  $V_5$ & 0.41725  	&  0.38017  		    \\ \hline 
  $V_7$ & 0.6585  	&  0.56026  		    \\ \hline 
\end{tabular}
\caption{ Two body pesudopotentials from the mapping $MF_{\langle c^{\dagger}c^{\dagger}cc\rangle}$ of the longer range four body interaction ($V_8=1,V_6=0$). The correlation $\langle c^{\dagger}c^{\dagger}cc\rangle$ are taken from the lowest energy $L=0$ state of the same four body interaction.}\label{tab:cccc8}
\end{table}

Note that since the correlations used to reduce the interactions in the two cases (Table \ref{tab:cccc6} and Table \ref{tab:cccc8}) are not the same, linearity property (which can be applied in the previous two cases $MF_{\langle c^\dagger c\rangle}$ and  $MF^2_{\langle c^\dagger c\rangle}$) does not apply here.

\begin{figure}  
  \begin{center}      
    \scalebox{0.68}{\input{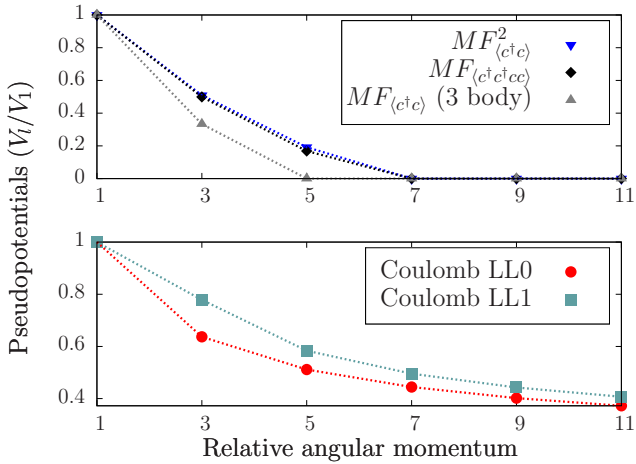}} 
\caption{Comparison of two body pseudopotentials for: Coulomb interaction, $MF^2_{\langle c^{\dagger} c \rangle}$ and $MF_{\langle c^{\dagger} c^{\dagger} c c \rangle}$ obtained from four body short-range repulsion ($2Q=27$), thermodynamical limit of pseudopotentials obtained from the  mean field approximation $MF_{\langle c^{\dagger} c \rangle}$ of three body short-range repulsion.} \label{ps_comp} 
                        
  \end{center} 
\end{figure}  

\subsection{Comparison of pseudopotentials}

Figure \ref{ps_comp} shows a comparison of the 
two body pseudopotentials obtained from mean field approximation of short-range four body repulsion, Coulomb repulsion and the two body interaction obtained from mean field approximation of short-range three body repulsion (described in Ref  \onlinecite{Sreejith17}) all normalized such that $V_1^{(2)}=1$.

\section{Mean field approximation and particle-hole anti-/symmetrization} \label{sec:PHSym}
In this section, we will explore the connection between the mean field approximation and the particle-hole symmetrization and antisymmetrization of multi body interactions. 
It was found in Ref [\onlinecite{Sreejith17}], that the spectrum of the mean field approximation of the short range three body interaction match exactly with the spectra of the interaction obtained by particle hole symmetrizing the short range three body interaction,\cite{Peterson08} suggesting that the two methods result in the same interaction. We will show here that the mean field approximation to a \emph{general} three body Hamiltonian is identical (upto additive constant chemical potential terms and overall scaling factors) to the particle hole symmetrization of the same.  Algebra involved being the same as in Wick's theorem, we can immediately generalize the ideas to the case of four body interactions.	

\subsection{Symmetrization of the three body interactions and its mean field approximation}
A general three body interaction can be written as 
\begin{equation}
{\mathcal H}^{(3)}=\frac{1}{3!3!}\sum_{{\bf p},{\bf q}}V_{{\bf p};{\bf q}}c_{p_{3}}^{\dagger}c_{p_{2}}^{\dagger}c_{p_{1}}^{\dagger}c_{q_{1}}c_{q_{2}}c_{q_{3}},
\end{equation}
where ${\bf p}\equiv (p_1,p_2,p_3)$, ${\bf q}\equiv (q_1,q_2,q_3)$ and the sum is over $-Q\leq p_i,q_i\leq Q $ without any constraints on the ordering inside ${\bf p}$.
\[
V_{\bf{p},\bf{q}}=\left\langle p_{1}p_{2}p_{3}\left|\mathcal{H}^{(3)}\right|q_{1}q_{2}q_{3}\right\rangle,
\]
The particle hole conjugation of the interaction is given by 
\begin{equation}
\overline{\mathcal{H}}^{(3)}=\frac{1}{3!3!}\sum_{{\bf p},{\bf q}}V_{{\bf p},{\bf q}} c_{p_{3}}c_{p_{2}}c_{p_{1}}c_{q_{1}}^{\dagger}c_{q_{2}}^{\dagger}c_{q_{3}}^{\dagger}.\label{phconjugate3}
\end{equation}

Shifting the creation operator to the right using the commutation relations reveals a relation between ${\mathcal{H}}^{(3)}$ and its particle hole conjugate.
\begin{equation}
\overline{\mathcal{H}}^{(3)}=\frac{1}{3!3!}\sum_{\bf{p},\bf{q}} V_{{\bf{p},\bf{q}}}\left[C^{(0)}_{{\bf p},{\bf q}}+C^{(2)}_{{\bf p},{\bf q}}+C^{(4)}_{{\bf p},{\bf q}}\right]-
{\mathcal{H}}^{(3)}.
\label{eq:V+V}
\end{equation}
Here
\begin{eqnarray}
&C^{(0)}_{{\bf p},{\bf q}}& = \frac{1}{3!}\sum_{Q,P\in S_3} (-1)^{PQ} \delta_{Q(q_1)P(p_1)} \delta_{Q(q_2)P(p_2)} \delta_{Q(q_3)P(p_3)}\nonumber\\
&C^{(2)}_{{\bf p},{\bf q}}& = -\frac{1}{2!}\sum_{Q,P\in S_3} (-1)^{PQ} \delta_{Q(q_1)P(p_1)} \delta_{Q(q_2)P(p_2)} c^\dagger_{Q(q_3)}c_{P(p_3)}\nonumber\\
&C^{(4)}_{{\bf p},{\bf q}}& = \frac{1}{2!2!}\sum_{P,Q\in S_3} (-1)^{PQ} \delta_{Q(q_1)P(p_1)} c^\dagger_{Q(q_3)} c^\dagger_{Q(q_2)} c_{P(p_2)} c_{P(p_3)}\nonumber
\end{eqnarray}
where $S_3$ is the permutation group over three objects.
These are precisely the terms that arise when Wick's theorem is used to relate the particle hole conjugate interaction (Eq \ref{phconjugate3}) to the normal ordered form, with the contraction being equivalent to setting $\langle c_i^\dagger c_j \rangle$ to be $\delta_{ij}$.

The first term $C^{(0)}$ gives a constant contribution to the right hand side of Eq \ref{eq:V+V}. The second term arising from $C^{(2)}$ is non zero only when a pair of entries in $\bf p$ match with a pair in $\bf q$. Considering that $V_{\bf{p},\bf{q}}$ is non zero only when $\sum p_i$ match with $\sum q_i$, we find that the $C^{(2)}$ is proportional to 
\begin{equation}
\sum_{p=-Q}^Q A_{pp} c_{p}^\dagger c_{p} \nonumber
\end{equation}
where
\begin{equation}
A_{pp}=\sum_{p_1,q_1,p_2,q_2=-Q}^Q V_{(p_1,p_2,p);(q_1,q_2,p)}\delta_{p_1q_1} \delta_{p_2q_2}\nonumber
\end{equation}
It can be seen that $A_{pp}$ is independent of $p$. Rotational symmetry of the interaction implies that the elements $V_{{\bf p},{\bf q}}$ are linear combintations of projectors onto fixed angular momentum subspaces ie 
\begin{equation}
V_{{\bf p},{\bf q}} = \sum_{L} a_L \left[P^{(3)}_L\right]_{{\bf p},{\bf q}} 
\nonumber
\end{equation}
Therefore $V_{{\bf p},{\bf q}}$ is a rotationally invariant tensor, i.e. invariant under the rotation $R$ (written in the $2Q+1$ dimensional representation).
\begin{equation}
V_{{\bf {p}},{\bf {q}}}=\sum_{{\bf \dot p},{\bf \dot q}} R_{p_1\dot{p}_1} R_{p_2\dot{p}_2} R_{p_3\dot{p}_3} V_{{\bf \dot{p}},{\bf \dot{q}}} {\bar{R}}_{\dot{q}_1 q_1} {\bar{R}}_{\dot{q}_2 q_2} {\bar{R}}_{\dot{q}_3 q_3}
\end{equation} 
Similarly $\delta_{p_1 q_1}\delta_{p_2 q_2}$ is a rotationally invariant tensor. So $A_{p_3q_3}$ obtained by contracting the four indices $p_1,p_2,q_1,q_2$ of the two tensors is also symmetric ie $A_{pq}=\sum_{\dot{p}\dot{q}} R_{p\dot{p}} \bar{R}_{\dot{q}q}A_{\dot{p} \dot{q}}$. Since there exists some rotation $R$ which takes an angular momentum $p$ to another angular momentum $p'$, we have that $A_{pp}=A_{p'p'}$. This implies that the term $C^{(2)}$ is simply a uniform chemical potential shift.

Finally, the term $C^{(4)}$ can be shown to be proportional to the mean field approximation of the three body interaction. Therefore we have that upto multiplicative and additive constants, 
\begin{equation}
\mathcal{H}^{(3)} + \overline{\mathcal{H}}^{(3)}  \propto MF_{\left \langle c^\dagger c\right \rangle } (\mathcal{H}^{(3)})
\end{equation}
Therefore the spectra of the particle hole symmetrization of the three body Hamiltonian,\cite{PetersonScarola19,Peterson08} and the mean field approximation of the same three body Hamiltonians are identical.\cite{Sreejith17}

\subsection{Particle Hole antisymmetrization of the four body interaction}

In this section, we ask if the relation shown in the previous section in the context of three body interactions generalize to the context of the four body interaction and show that such a simple exact relation does not exist between the particle hole symmetrization and the mean field Hamitlonians. Consider the expansion of the particle hole conjugate of the four body interaction in terms of a sequence of normal ordered operators
\begin{equation}
\overline{\mathcal{H}}^{(4)} = \mathcal{H}^{(4)} - \mathcal{H}^{(4\to 3)} + \mathcal{H}^{(4\to 2)} - \mathcal{H}^{(4\to 1)} + \mathcal{H}^{(4\to 0)}
\end{equation}
where the first term is the four body interaction and the terms $\mathcal{H}^{(4\to n)}$ for $n=3,2,1,0$ are obtained under sequence of application of $MF_{\left\langle c^\dagger c\right \rangle}$. Equivalently these are the terms obtained after $1$, $2$, $3$ and $4$ contractions $\left\langle c^\dagger c\right \rangle\propto \delta_{ij}$. $ \mathcal{H}^{(4\to 0)}$ is a constant shift. As discussed in the last section rotational invariance implies that the $\mathcal{H}^{(4\to 1)}$ is also a constant chemical potential shift.

The above expression tells us that unlike the case of the three body interaction, it is the particle hole-antisymmetrization of the four body interaction that contains fewer-body interactions. Upto constant shifts, we get the following results
\begin{eqnarray}
&{\mathcal{H}}^{(4)}-\overline{\mathcal{H}}^{(4)}  = {\mathcal{H}}^{(4\to 3)} - {\mathcal{H}}^{(4\to 2)}\nonumber\\
&{\mathcal{H}}^{(4)}+\overline{\mathcal{H}}^{(4)}  = 2{\mathcal{H}}^{(4)}- {\mathcal{H}}^{(4\to 3)} + {\mathcal{H}}^{(4\to 2)}\nonumber
\end{eqnarray}
where ${\mathcal{H}}^{(4\to 3)}\propto MF_{\left\langle c^\dagger c\right \rangle}(\mathcal{H}^{(4)})$ and ${\mathcal{H}}^{(4\to 2)}\propto MF^2_{\left\langle c^\dagger c\right \rangle}(\mathcal{H}^{(4)})$. Neither particle hole symmetrization nor antisymmetrization produces a simple interaction that can be expanded in terms of positive pseudopotentials. In general, for (even)odd $n$, the particle hole (anti)symmetrization of the $n$-body interaction produces an interaction that can be interpretted as a sum of $n-1$ and fewer body interaction albeit with some negative pseudopotentials.

\section{Numerical tests of the mean field approximations}
\label{sec:numtests}
In this section, we present the results of numerical tests of the three mean field approximations described in Sec \ref{MF:desc}. In particular we focus on the states at filling fraction $\nu=3/5$ at which the short range four body interaction with $V_6=1$ (other pseudopotentials are zero) produces an incompressible ground state. We compare the spectra of the approximations with that of the original four body interaction. Since the mean field approximations produce the same two body interaction obtained in Ref [\onlinecite{Kusmierz18a}], the numerical tests given below extend the numerical tests presented there.

\subsection{Hamiltonian spectra}

\begin{figure}  
\begin{center}      
\scalebox{0.68}{\input{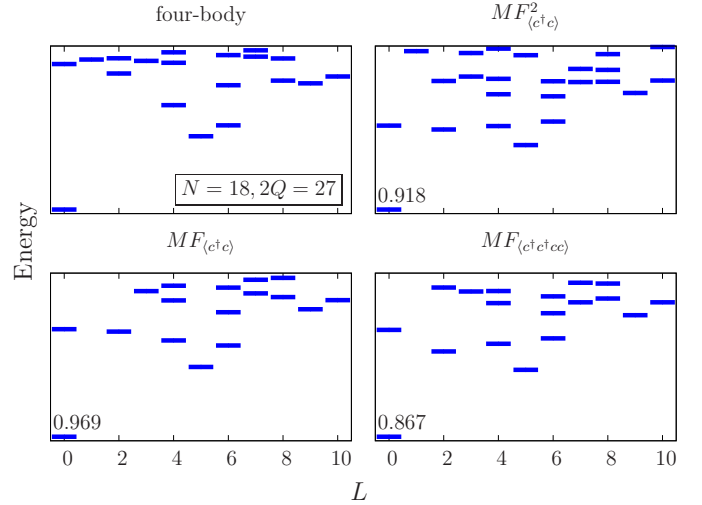}} 
\caption{The spectra of a four body short-range repulsion Hamiltonian and its mean field approximations at a filing factor $\nu = 3/5$, $N=18, 2Q = 27$. The numbers next to the ground states show the overlap of the corresponding state with the ground state of the four body interaction.} \label{spev6} 
\end{center} 
\end{figure}

An incompressible ground state representing a filling fraction of $3/5$ is produced by the short range four body interaction in a system of $N$ (a multiple of 3) electrons on a sphere pierced by $2Q=5N/3-3$ radial flux quanta. This incompressible state corresponds to the $k=3$ Read-Rezayi state.\cite{Read99} Fig \ref{spev6} (top-left) shows the spectrum of such a system of $N=18$ particles.
This incompressible state for a system of $N$ particles can be written as\cite{Read99}
\begin{multline}
{\mathcal A} {\Big[}\Psi_{\frac{1}{3}}({\bf z}) \Psi_{\frac{1}{3}} ({\bf w}) \Psi_{\frac{1}{3}}({\bf r}) \times \\ \times \prod_{i,j=1}^{N/3}(z_i-r_j) \prod_{i,j=1}^{N/3}(z_i-w_j) \prod_{i,j=1}^{N/3}(r_i-w_j){\Big]}\label{tripartite}
\end{multline}
where ${\bf z, w, r}$ are partitions into three equal parts of the $N$ coordinates. The function $\Psi_{\frac{1}{3}}$ is the Laughlin state at filling fraction $1/3$. The symbol ${\mathcal A}$ indicates antisymetrization over $N$ coordinates and ensures that the function represents a wavefunction of $N$ indistinguishable particles. The function is expressed in the language of disc geometry, but can be straightforwardly mapped to the spherical geometry using a stereographic projection. 

Just above the gapped ground state is a neutral mode whose wavefunction corresponds to the one in which one of the partitions $\Psi_{\frac{1}{3}}$ has a neutral excitation. \cite{Sreejith11a, Sreejith12a,ExcitonsCFKamila96} Using this construction, the allowed quantum numbers of the neutral mode can be inferred to be $0<L\leq N/3$.
In Fig \ref{spev6} (top-left), the neutral mode can be seen to extend up to an angular momentum $L=6$ as expected, however the mode merges into the bulk spectrum at low angular momenta. 

The spectrum of the two body interaction obtained using the mean field approximation $MF^2_{\langle c^\dagger c \rangle}$ is shown in the Fig \ref{spev6}(top right). The spectrum contains a unique $L=0$ ground state with a high overlap with the Read-Rezayi state. A mode of excitations can be seen above this whose counting matches at larger angular momentum but appears to differ at lower angular momenta. Note that for neutral excitations (which are made of a quasiparticle-quasihole pair), lower angular momenta correspond to states where the quasiparticle and quasihole are close to each other. Relative agreement in the spectra as angular momentum increases suggests that this mean field approximation reproduces the right long-distance physics. The three body interaction obtained from $MF_{\langle c^\dagger c \rangle}$ also produces an incompressible state (Fig \ref{spev6}(bottom left)) with a high overlap with the Read-Rezayi state as well as a neutral mode with the right quantum numbers. Spectra of the two body interaction obtained using $MF_{\langle c^\dagger c^\dagger c c \rangle}$, shown in Fig \ref{spev6}(bottom right) is qualitatively similar to that of the spectra of the two body interaction obtained from $MF^2_{\langle c^\dagger c \rangle}$. Note that the pseudopotentials depend on $2Q$. The calculations presented here use mean field psuedopotentials at the respective fluxes and not the ones inferred for the thermodynamic limt.

\begin{figure}  
  \begin{center}      
  \includegraphics[width=\columnwidth]{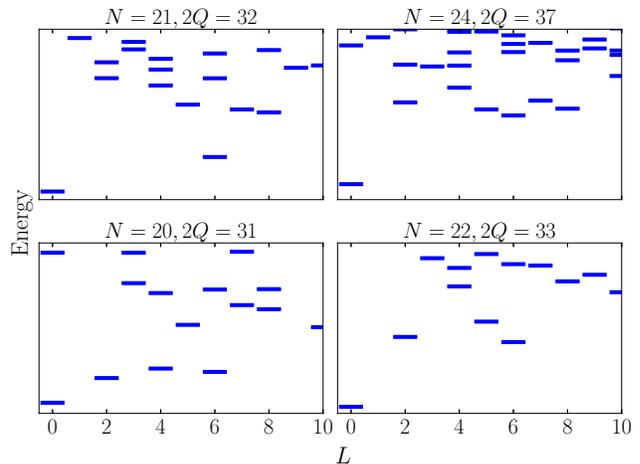}
\caption{The spectra of the mean field approximation $MF^2_{\langle c^\dagger c \rangle}$ to the short range four body Hamiltonian with nonzero pseudopotential for $V_6$ at $N=21, 2Q = 32$,  $N=24, 2Q=37$; $N=20, 2Q = 31$ and $N=22, 2Q = 33$.}\label{MFonly}
  \end{center}  
\end{figure}

The results presented in Fig \ref{spev6} are for the largest system in which all interactions were studied. Though diagonalization of four body interaction in larger systems is not easy, the quantum numbers of the low energy states can be inferred from the trial wavefunction approach discussed above. The mean field two body interaction can be diagonalized in bigger systems and the low energy quantum numbers can be compared with those from the trial wavefunctions. Fig \ref{MFonly} (top left) shows the spectrum of the two body interaction obtained using $MF^2_{\langle c^\dagger c\rangle}$ in a system $N=21,2Q=32$. The interaction again produces a homogeneous incompressible ground state and a neutral mode. However we find that the neutral mode ends at angular momentum $L=8$ instead of $L=7$. The top right panel shows the spectrum of the mean field two body interaction in a larger system $N=24,2Q=37$. Here we find that the neutral mode counting matches with the predicted angular momentum of $L=8$.

\begin{figure}  
  \begin{center}      
    \scalebox{0.68}{\input{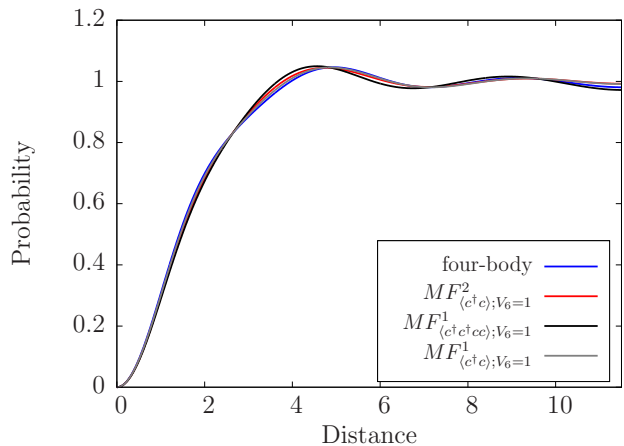}}
\caption{Pair correlation functions calculated from the incompressible ground state at flux $2Q=5N/3-3$ of the short range four body interaction as well as its mean field approximations.}\label{fig:cor_f1}     
  \end{center} 
\end{figure}

The structure of the neutral mode is indirectly encoded in the ground state pair correlation functions\cite{PhysRevB.33.2481} and therefore we expect that these functions should also be reproduced by the approximate Hamiltonians. Figure \ref{fig:cor_f1} shows the pair correlation functions in the incompressible ground state of the four body interaction as well as in the ground states of the different approximate Hamiltonians shown in Fig \ref{spev6}.

\begin{figure}  
\begin{center}      
\scalebox{0.65}{\input{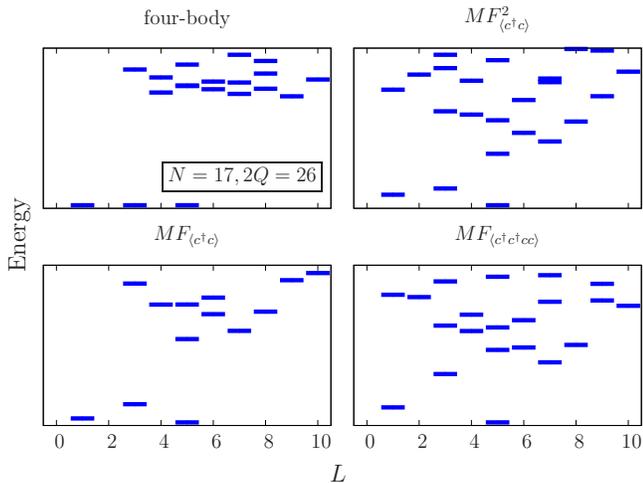}}
\caption{The spectra of short range four body repulsion ($V_6=1,V_8=0$) and its mean field approximation at $N=17, 2Q = 26$ (corresponding to two quasihole states).}\label{17_26}
\end{center} 
\end{figure}

Figure \ref{17_26} shows the spectrum of a system $N=17,2Q=26$ which has one electron and a flux less than that in the incompressible state. The low energy spectrum arises from two quasiholes of the Read-Rezayi state. The wavefunctions at low energies can be understood in terms of a three-partition structure similar to that in Eq \ref{tripartite}, wherein the three partitions now contain $6,6$ and $5$ electrons in the states $\Psi_{\frac{1}{3}}$, $\Psi_{\frac{1}{3}}$ and $\Psi_{\frac{1}{3}}^{\rm 2qh}$; the two quasiholes correspond to two quasiholes in the Laughlin state in one of the three partitions.\cite{Sreejith12a} Using this structure of the wavefunction, the angular momenta of the low energy states can be calculated to be $L=1,3,5$ and can be verified in the exact spectrum. Here we find that the spectra of all mean field approximations closely resemble the spectrum of the four body interaction. 	

Fig \ref{MFonly} (bottom left) shows the spectrum of the system $N=20,2Q=31$ which is again obtained by removing an electron and a flux from the incompressible system at $N=21,2Q=32$. This can again be understood as a similar two quasihole state and the angular momenta of the low energy states can be calculated to be $L=0,2,4,6$. This exactly matches what is seen the spectra of the mean field two body Hamiltonian.

\begin{figure}  
  \begin{center}      
    \scalebox{0.65}{\input{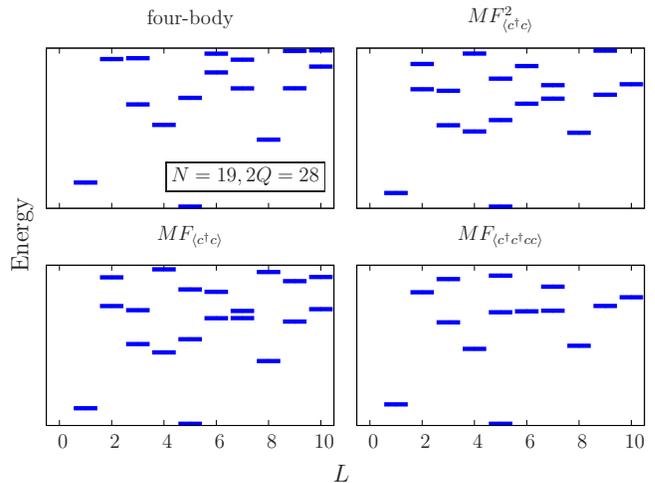}}
\caption{The spectra of short range four body repulsion ($V_6=1,V_8=0$) for a system of $N=19, 2Q = 28$ (corresponding to two quasiparticles).}\label{19_28}                                 	  \end{center} 
\end{figure}

Fig \ref{19_28} shows the spectra for a system at $N=19,2Q=28$, with one electron and one flux more than in the incompressible state. The low energy spectrum is expected to correspond to a system containing a pair of quasiparticles. The wavefunction can be understood as containing three partitions (similar to Eq-\ref{tripartite}) of $6,6$ and $7$ particles in the states $\Psi_{\frac{1}{3}}$, $\Psi_{\frac{1}{3}}$ and $\Psi_{\frac{1}{3}}^{\rm 2qp}$; the two quasiparticles being in the last partition. The quantum numbers of the two quasiparticle state, within this picture is then the same as the quantum numbers $L=1,3,5$ of the two quasiparticles of Laughlin state in the last partition. A clearly separated quasiparticle branch with this counting cannot be seen even in the original four body interaction. Since the two quasiparticles are closer to each other with higher probability in the higher angular momentum states, it is expected that such a counting based on trial wavefunctions should work only in the low angular momentum limit. 	
The spectra (Fig \ref{19_28}) of all the three mean field approximations match with what is seen in the actual spectrum of the four body interaction. 
Fig \ref{MFonly}(bottom right) shows the spectra of the mean field two body interaction in the next bigger system where we expect a two quasiparticle state. Based on the wavefunctions described above, the quantum numbers of the low energy states are expected to be $L=0,2,4,6$. The quantum numbers in the spectra match with this in the low angular momentum limit.

\begin{figure}  
  \begin{center}      
    \scalebox{0.65}{\input{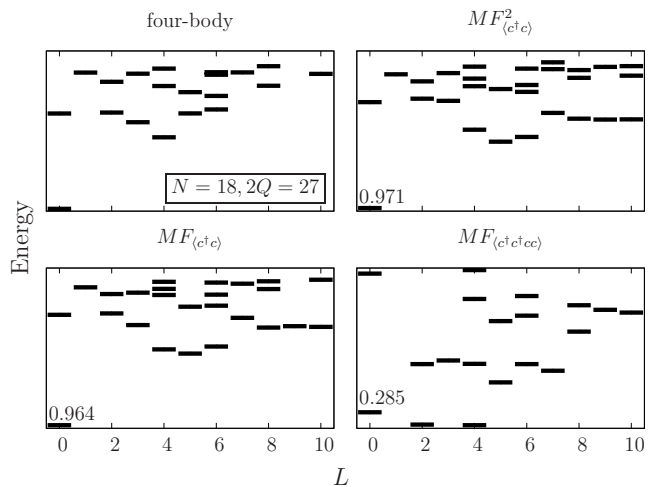}}    
\caption{The spectra of the longer range four body repulsion Hamiltonian with nonzero pseudopotential for $V_8=1$ alone and its mean field spectra in a system of size $N=18, 2Q = 27$.}\label{spev8}                           	  \end{center} 
\end{figure}

For completeness, we also explore the spectra of the mean field approximation to the longer range $V_8=1,V_6=0$ interaction. In general, the interaction does not produce an incompressible ground state in the same flux sector $2Q=\frac{5N}{3}-3$ that we have studied. A gapped homogeneous state is produced in the specific case of $N=18$ but not in $N=15$. Results for the spectrum of this interaction at $N=18$ is shown in Fig \ref{spev8}. In this case, the spectrum of the four body interaction is closely reproduced by the the $MF_{\langle c^\dagger c \rangle }$ and $MF_{\langle c^\dagger c \rangle }^2$ approximations but not by $MF_{\langle c^\dagger c^\dagger c c\rangle }$. The difference in the spectrum is not surprising given that the two body interaction obtained in the two methods (Table \ref{tab:cccc8} and \ref{tab:mfcc}) appear qualitatively different. For the case of $N=15$ where there is no clear gap in the spectrum, we can still construct the approximate Hamiltonian using the correlation function in the ground state of the $L=0$ sector. In this case we find that the $MF_{\langle c^\dagger c^\dagger c c\rangle }$ produces a spectrum closer to the four body interaction.

\section{Conclusions}
We have presented three approaches for approximating the four body interaction to obtain fewer body Hamiltonians and tested the approach on systems around the flux value $2Q=\frac{5}{3}N-3$, where the short ranged four body interaction produces a gapped ground state. Evidence from numerical diagonalization  of finite size systems suggest that the approximation schemes produce a good effective model of the physics of at the $3/5$. The two body pseudopotentials for approximation can be estimated to be close to $V_1:V_3:V_5=6:3:1$. Comparison with previous studies in Ref \onlinecite{Kusmierz18a} suggest that the obtained two body approximations are indeed the optimal two body interactions that produce the Read Rezayi state. Similar approximation to the short range three body interaction produces the corresponding optimal interaction that approximates the low energy physics around the Pfaffian state. The mean field approximation of three body interaction is exactly same upto constants as its particle hole symmetrization, but such a relation is not true for the $n>3$ body interactions. The two body mean field approximations, which seems to accurately reproduce the spectra, form only a part of the particle hole symmetrization of of the four body interaction. It will be interesting to explore the importance of the particle hole symmetry breaking and the symmetry preserving corrections  to the mean field approximations.

 \section*{ACKNOWLEDGMENTS}

The authors (BK, AW) acknowledge financial support from the Polish NCN Grant No. 2014/14/A/ST3/00654. SGJ thanks Yuhe Zhang and J K Jain for useful discussions.

\bibliography{4body}
\bibliographystyle{apsrev}

\appendix

\section{Fitting functions of approximate pseudopotentials }\label{app1}

\begin{table}[!h]
\centering
\begin{tabular}{|c||c|c|c|}
\hline
 \multicolumn{4}{|c|} {$MF^2_{\langle c^\dagger c \rangle}(V^{(4)}_6=1,V^{(4)}_8=0)$ } \\ \hline
$V^{(2)}_l$ & $a$ & $b$ & $c$  \\
\hline
  $l=1$ & 7.24975   &  -14.0437         & 0.659577      \\ \hline
  $l=3$ & 3.50013   &  -1.88354         & 1.42402   	\\ \hline
  $l=5$ & 1.24996  	&  0.939755  		& 4.45989     \\ \hline  \hline 
\multicolumn{4}{|c|} {$MF^2_{\langle c^\dagger c \rangle}(V^{(4)}_6=0,V^{(4)}_8=1)$ } \\ \hline
$V^{(2)}_l$ & $a$ & $b$ & $c$  \\
\hline
  $l=1$ & 6.50009   &  -21.3186         & 0.608319    	\\ \hline
  $l=3$ & 2.75182   &  -4.0066         	& 0.552965       \\ \hline
  $l=5$ & 1.00101   &  2.62568        	& 4.58301       \\ \hline
  $l=7$ & 1.75123  	&  3.43484  		& 6.05555     	\\ \hline  \hline
 \multicolumn{4}{|c|} {$MF_{\langle c^\dagger c \rangle}(V^{(4)}_6=1,V^{(4)}_8=0)$ } \\ \hline
$V_l^{(3)}$ & $a$ & $b$ & $c$  \\
\hline
  $l=3$ & 3.16047   &  -2.63259         & 1.71567   	\\ \hline
  $l=5$ & 1.18517  &  0.494979         	& 3.46206       \\ \hline
  $l=6$ & 0.987714 &  -0.085466        	& 4.79941        \\ \hline  \hline

   \multicolumn{4}{|c|} {$MF_{\langle c^\dagger c \rangle}(V^{(4)}_6=0,V^{(4)}_8=1)$ } \\ \hline
$V_l^{(3)}$ & $a$ & $b$ & $c$  \\
\hline
  $l=3$ & 2.10724   &  -5.28236         & 1.44103    	\\ \hline
  $l=5$ & 0.658232  &  0.7258         	& 1.51154       \\ \hline
  $l=6$ & 0.0652811 &  0.538105        	& 5.42847        \\ \hline
  $l=7$ & 1.57983   &  1.8651       	& 3.92187       \\ \hline
  $l=8$ & 0.922529 	&  -0.64921  		& 5.68722    	\\ \hline 
   
\end{tabular}
\caption{The coefficients of a function $V_{l}^{(n)} (2Q) = a+\frac{b}{c-2Q}$, which is a best fit of $n=2$ and $n=3$ body pseudopotentials obtained from mean field approximation of four body interaction with a single nonzero four body pseudopotential ($V_6$ or $V_8$). 
}
\label{tab:coef}
\end{table}

\section{Expansion of $\langle c_{p_1}^\dagger c_{p_2}^\dagger  c_{q_2} c_{q_1}\rangle$ in angular momentum channels}
\label{App:4fermioncorrel}

\begin{figure}[h!]
\includegraphics[width=\columnwidth]{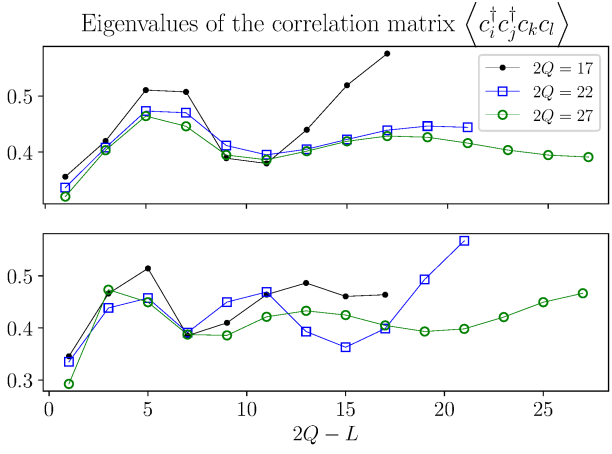}
\caption{Eigenvalues of the correlation matrix $M_{\bf{pq}}=\langle c_{p_1}^\dagger c_{p_2}^\dagger  c_{q_2} c_{q_1}\rangle$ plotted as a function of the total angular momentum quantum number of the eigenvector. Top figure shows the correlation in the ground state of the short range four body interaction $V_8=0,V_6=1$ and the bottom figure shows the same for ground state of the $L=0$ sector ground state of the longer range interaction $V_8=1,V_6=0$. }\label{Corelations4pt}	\end{figure}

\begin{figure}[h!]
\includegraphics[width=\columnwidth]{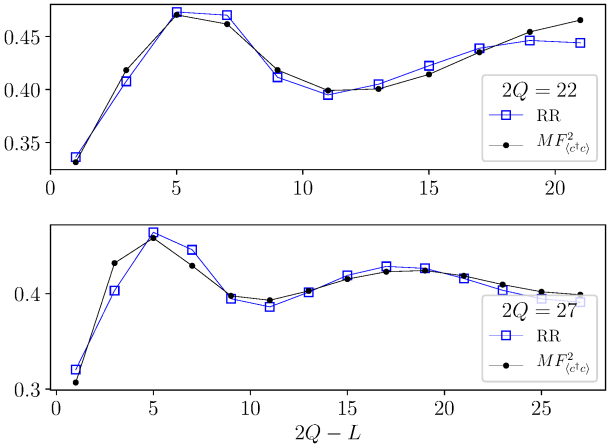}
\caption{Plot shows the eigenvalues of the correlation matrix as a function of the angular momentum similar to Fig \ref{Corelations4pt}. The figure compares the correlations in the Read-Rezayi state with that in the ground state of the $MF^2_{\langle c^\dagger c \rangle }$ approximation of short range interaction in two different system sizes $N=15,2Q=22$ (top) and $N=18,2Q=27$(bottom)}	\end{figure}

\end{document}